\documentclass[prd,twocolumn,superscriptaddress,nofootinbib]{revtex4-2}
\usepackage{natbib}
\usepackage{graphicx}
\usepackage{bm}
\usepackage{amssymb}
\usepackage{color}
\usepackage{amsmath}
\usepackage{mathrsfs}
\usepackage{amstext}
\usepackage[english]{babel}
\usepackage{latexsym}
\usepackage{array}             
\usepackage{multirow}         
\usepackage[usenames,dvipsnames]{xcolor}
\usepackage[colorlinks=true,citecolor=Blue,linkcolor=RubineRed,urlcolor=Blue]{hyperref}
\usepackage{tocloft}
\usepackage{multibib}

\newcommand{\beq}{\begin{equation}}
\newcommand{\eeq}{\end{equation}}
\newcommand{\beqa}{\begin{eqnarray}}
\newcommand{\eeqa}{\end{eqnarray}}

\begin{document}


\title{Pattern formation  from  Gauge/Gravity Duality}

\author{Chuan-Yin Xia}
\affiliation{Center for Gravitation and Cosmology, College of Physical Science and Technology, Yangzhou University, Yangzhou 225009, China}
\affiliation{Center for Theoretical Physics , Hainan University, Haikou 570228, China}

\author{Hua-Bi Zeng}
\affiliation{Center for Theoretical Physics , Hainan University, Haikou 570228, China}
\affiliation{Center for Gravitation and Cosmology, College of Physical Science and Technology, Yangzhou University, Yangzhou 225009, China}

\begin{abstract}
In the framework of the AdS/CFT correspondence, we find a neutral complex scalar field dynamics in  a $2+1$ dimensional black hole background which can provide a scheme for studying the pattern formation process in $1+1$ dimensional reaction-diffusion systems. The patterns include plane  wave,
defect turbulence, phase turbulence, spatio-temporal intermittency where defect chaos coexists with stable plane  wave, and coherent
structures. A phase diagram is obtained by studying the linear instability of the plane wave
solutions to determine the onset of the holographic version of the BFN instability. Near the critical temperature the holographic model is dual to the one-dimensional complex Ginzburg-Landau equation (CGLE), which has been studied extensively in reaction-diffusion systems. While at low temperature the holographic theory is different from CGLE. 
\end{abstract}

\maketitle

\textit{Motivation--}
As a remarkable result emerged from
string theory, the duality between a classical gravity theory and a quantum field theory  living on its boundary  is called AdS/CFT correspondence, also known as Gauge/Gravity duality or holography\cite{Maldacena1999,Gubser1998,witten1998}.
The correspondence provides an unique method for studying a strongly coupled quantum many-body system in equilibrium \cite{zaanen2015,ammon2015,hartnoll2018,zaanen2021} or out of equilibrium \cite{Liu2020},  has shown great power and potential in condensed matter physics (AdS/CMT). Recently,
most holographic non-equilibrium
applications focus on extending the holographic superconductor model \cite{Gubser2008,Hartnoll2008,Herzog2009} out of equilibrium \cite{zhao2024dynamical,Guo2020,zeng2018,doi:10.1126/science.1233529,PhysRevLett.127.101601,PhysRevLett.131.221602,PhysRevLett.110.015301}.
Here we firstly exhibit that the holographic duality can also provide a dual gravitational description of non-equilibrium pattern formation dynamics.

 Understanding the mechanism of spontaneous pattern formation out of equilibrium in fluids, plasmas, cosmology, crystals solidifying from a melt, and so on is one of the fundamental questions in nonequilibrium physics\cite{cross_greenside_2009,RevModPhys.65.851,gollub1999pattern,liddle_lyth_2000,Gollub1999,1990IJMPB}.
Rather than a physical system, pattern formation is also frequently
observed in a chemistry or biology system\cite{maini1997spatial,petrov1997resonant,1952turing}.
In contrast to pattern formation within thermodynamic
equilibrium which is rooted in the minimization of (free) energy,
patterns emerging in nonequilibrium systems can only be
understood within a dynamical framework, even if the
patterns of interest are time independent.
More often than not, when a system is driven far from equilibrium, spatially uniform structures become unstable toward the growth of small perturbations, which leads to dynamics that amplify fluctuations and increase complexity.
Late-time dynamics is dominated by the fastest-growing fluctuating modes, whose characteristic length and time scales determine the resulting spatiotemporal patterns, eventually stabilized by nonlinear and dissipative mechanisms\cite{RevModPhys.65.851}.
In such a dynamical framework, dynamical instabilities and nonlinear mode coupling mechanisms are crucial for pattern formation \cite{gollub1999pattern}.
A typical nonlinear equation that have been most used in pattern formation is the  cubic complex Ginzburg-Landau equation 
(CGLE) \cite{RevModPhys.65.851,Bekki1985133,Shraiman1992241,Chate_1994,19841581,VANSAARLOOS1992303,Brusch2000,Brusch20011,hohenberg2015}
\begin{equation}
\label{CGLE}
\partial_t\phi=(1+i\alpha)\nabla^2\phi +\phi-(1+i\beta)|\phi|^2\phi,
\end{equation}
 since it describes the general dynamical characteristics of an extended system close to a Hopf bifurcation\cite{RevModPhys7499}. It  was first derived in the studies of Poiseuille flow \cite{Stewartson_Stuart_1971} and describes a variety of physical phenomena at a qualitative or even quantitative level, from nonlinear waves to second-order phase transitions, from superconductivity, superfluidity, Bose-Einstein condensation to liquid crystal strings in field theory.
Formally, it is a semi-parabolic nonlinear partial differential equation that can describe a single-component reaction-diffusion system, also actual chemical systems \cite{kolmogorov1937study,Kuramoto1984,Nicolis_1995}. 
Where $\phi=\phi(\vec{r},t)$ is a complex field to describe spatio-temporal phenomena in continuous media, in a chemical
system $\phi$ is the concentration.
$(1+i\alpha)\nabla^2\phi$ and $\phi-(1+i\beta)|\phi|^2\phi$ are the diffusion term and reaction term respectively, and the real parameters $\alpha$ and $\beta$ can both determine the properties of the pattern, in principle can  be decided from experiments \cite{PhysRevE1993}.
The simplest solutions of Eq. \eqref{CGLE} are plane wave solutions
    $\phi=\phi_a e^{i q x+i \omega t}$,
where
\begin{equation}
\label{GLplane_1}
\phi_a^2 = 1-q^2, \quad \omega = -\alpha q^2 - \beta \phi_a^2.
\end{equation}
The dependence of the wave’s frequency $\omega$ on the wavenumber $q$  illustrates a nonlinear dispersion of the CGLE \eqref{CGLE}.
To investigate the stability of the plane wave solution when $q=0$, the perturbation should be introduced in the way of
\begin{equation}
\label{GLperturbation}
\phi=(\phi_a+\delta_{1} e^{\lambda t+ikx}+\delta^*_{2} e^{\lambda^* t-ikx}) e^{i \omega t},
\end{equation} 
where the scalars $\delta_{1}$ and $\delta^*_{2}$
 denote the amplitudes of the small perturbations.
Then substituting Eq. \eqref{GLperturbation} into Eq. \eqref{CGLE}, one can get  eigenvalue equation about $\lambda$ which only dependent on the parameters $\alpha$ and $\beta$. 
As long as the real part of $\lambda$ is smaller than zero,  the so-called BFN instability \cite{Benjamin1,Benjamin2} criterion 
\begin{equation}
\label{glBF}
1+\alpha\beta>0
\end{equation}
can be obtained (See appendix \ref{appA} for a detailed review). 
The outer boundary of the BFN instability is called the BFN line. Above the line in the plane wave solution is  unstable and move to Spatiotemporal chaos solution  through a supercritical Hopf bifuration \cite{Shraiman1992241}.
Below the BFN line, it can exist the spatio-temporal intermittency where defect chaos coexists with stable plane  wave   and coherent structures appear \cite{Chate_1994,19841581,VANSAARLOOS1992303}. These structures are
related to experiments in Rayleigh-Bernard convection,
hydrothermal nonlinear wave, chemical systems and so on, 
which is as reviewed completely in \cite{RevModPhys7499}.

When $\alpha=\beta=0$, the CGLE Eq. (\ref{CGLE}) degenerates into the Ginzburg-Landau equation (GLE) in the symmetry broken phase,  the neutral scalar $\phi$ is the order parameter of the second phase transition. From the application of Gauge/Gravity duality we found  there is a gravitation model dual to 
the Ginzburg-Landau(GL) phase transition theory near the critical point with also a neutral scalar field as the order parameter, proposed in \cite{Iqbal2010}, 
based on the theory  we successfully constructed a holographic model by introducing two parameters in the bulk theory similar to CGLE, the  dynamics of the neutral scalar field living in a charged black hole can also demonstrate immense kinds of pattern formation on the boundary that arise naturally and autonomously from a spatial homogeneous uniform oscillating state.

\textit{Model from holography--}
Following  \cite{Iqbal2010},   we consider a ($d+1)$-dimensional anti–de Sitter (AdS$_{d+1}$) spacetime, the Reissner-Nordstr\"om (RN) black hole background with a neutral complex scalar field $\Psi$ living from the horizon of the black hole to infinity.
The RN black hole is a solution of the Einstein-Maxwell theory with negative cosmological constant $\Lambda = - d  (d - 1)/2 \ell^2$,
\begin{equation}
\mathcal{L} = R - 2 \Lambda + \alpha F_{\mu\nu} F^{\mu\nu}.
\end{equation}
We further focus on the $d = 2$ case to study the pattern formation dynamics of a 1+1 dimensional system living on the boundary of the $AdS_3$ RN black hole, where
\begin{equation}
A_t = - \mu \ln z, \quad f = 1 - z^2 + \frac{\mu^2}{2} z^2 \ln z,
\end{equation}
with temperature
\begin{equation}
T = \frac1{4 \pi} \left( 2 - \frac{\mu^2}{2}\right).
\end{equation}
In the Eddington coordinate, $dt \to dt + du/f$, the metric has the form
\begin{equation} \label{dsE}
ds^2 = \frac{\ell^2}{z^2} \left( -f(z) dt^2 - 2 dtdz + dx^2 \right).
\end{equation}

In the background of the RN black hole, we
consider a neutral scalar with it's Lagrangian reads
\begin{equation}
\mathcal{L}_\Psi =\frac{1}{2 \kappa \lambda} ( - V_M(\Psi) -V_K(\Psi)).
\end{equation}
One term of the  the Lagrangian is  the nonlinear Mexican hat potential
\begin{equation}
V_M = \frac{1}{4 \ell^2} ((1+i \beta )\Psi^2\Psi^{*2} +2\Psi\Psi^*m^2 \ell^2) ,
\end{equation}
and the other term is the kinetic energy term
\begin{equation}
V_K=  \frac{1}{2} (\partial_t \Psi\partial^t\Psi^* +\partial_z \Psi\partial^z \Psi^*+(1+i \alpha )\partial_x \Psi\partial^x \Psi^*),
\end{equation}
The equation of motion for the complex scalar field has the following form
\begin{eqnarray}
(2 z^2\partial_t\partial_z  - z \partial_t) \Psi=z^2(1+i \alpha ) \partial_x^2 \Psi 
-(1+i \beta ) \Psi ^2 \Psi ^*
\nonumber\\
+z^2 \partial_z(f \partial_z \Psi) -z f \partial_z \Psi-m^2 \Psi.
\label{EOM}
\end{eqnarray}
The asymptotic expansion of the field near the boundary is
\begin{equation}
\label{expansion}
\Psi\big|_{z=0} =\Psi^-  z^{\triangle^-}+ \Psi^+ z^{\triangle^+} ,
\end{equation}
where
\begin{equation}
\triangle^\pm = \frac{2 \pm \sqrt{4 + 4 m^2}}{2}.
\end{equation}
Standard quantization is adopted on the boundary, where $\Psi^-$ can be regarded as the source of the operator in the boundary field theory and $\Psi^+$ can be regarded as the expected value of the scalar operators $\mathcal{O}$. Setting the source of the operator $\Psi^-$=0, one obtains a spontaneous symmetry-breaking state in this holographic setting when the temperature of the black hole below a critical value. At the horizon $z=1$, $\Psi$ is regular \cite{Iqbal2010} and it is automatically satisfied, which can be found numerically. The two parameter $\alpha$ and $\beta$ are expected to play similar roles of the two parameter in CGLE Eq.(\ref{CGLE}). By setting $\alpha=\beta=0$, this model was firstly proposed and studied in $d=3$ \cite{Iqbal2010}, which duals to the GL second order phase model near the critical temperature. Recently the model was extended to the $d=2$
case which was found to  dual to a discrete version of GL theory  for structure phase transition in one spatial dimension\cite{Xia}. 

In this letter, we set the mass to $m^2=-0.99$, which is a little above the Breitenlohner-Freedman bound $-d^2/4(d=2)$,  the critical temperature is $T_c=0.014$, corresponding to $\mu_c=1.9$. We will show that by tuning on $\alpha$ and $\beta$ in the symmetry broken phase, the system will demonstrate various pattern formations as observed in a one-dimensional reactive-diffusion system.  We chose a typical symmetry broken state $T=0.29T_c$.
The size of a one-dimensional reaction-diffusion system growing on the boundary is set to $L=500$. In order to solve the  dynamic equation \eqref{EOM}, the following numerical methods are necessary the Chebyshev spectral method is used in the $z$ direction, the Fourier spectral method is used in the $x$ direction.  Specifically, the number of points in the $z$ direction and the $x$ direction is $n_z=30$ and $n_x=500$, respectively. The fourth-order Runge-Kutta method is used to simulate the evolution of the system in the time direction, and the time step is $h=0.01$.

\textit{Plane wave solution, holographic Benjamin–Feir instability and phase diagram--}
Similar to the CGLE we begin with the  plane wave solutions in the holographic model, which are related to the parameters $(\alpha,\beta,q)$ and take the form of
\begin{equation}
\Psi(z,t,x)=\Psi_a(z) e^{i \omega t+ i q x},
\label{planewave}
\end{equation}
where $\Psi_a(z)$ is a monotonically increasing function with respect to $z$. It should be emphasized that the solutions of the dynamic Eq. \eqref{EOM} are not always plane wave solutions unless the ansatz Eq. \eqref{planewave} is adopted. Then we can obtain the Fourier transform form of the dynamic Eq. \eqref{EOM} \begin{eqnarray}
i\omega(2 z^2\partial_z  - z ) \Psi_a=-q^2z^2(1+i \alpha )  \Psi_a -(1+i \beta ) \Psi_a ^2 \Psi_a ^*
\nonumber\\
+z^2 \partial_z(f \partial_z \Psi_a) -z f \partial_z \Psi_a-m^2 \Psi_a.
\label{EOMplane}
\end{eqnarray}
By the way, Eq. \eqref{EOMplane} can be easily  solved by the Newton-Raphson iteration method.
According to Eq. \eqref{expansion}, the order parameter of the boundary field theory reads $\mathcal{O}= \mathcal{O}_a e^{i \omega t+ i q x}$.
After solving the Eq. \eqref{EOMplane}, we find both $\mathcal{O}_a$ and $\omega$ are related to the parameters $(\alpha,\beta,q)$.
Specifically, they can be fitted by
\begin{equation}
\mathcal{O}_a^2=\mathcal{O}_{0}^2-(0.0276+0.00219\alpha\beta)q^2+0.000190\beta^2,
\end{equation}
and
\begin{equation}
\omega= -(0.108\alpha-0.127\beta)q^2 -5.02\beta\mathcal{O}_a^2,
\end{equation}
where $\mathcal{O}_{0}^2= 0.00373$ is defined at $q=\alpha=\beta=0$. All the plane wave solutions obtained from Eq. \eqref{EOMplane} can be verified by the dynamic Eq. \eqref{EOM}. 
Please note that the plane wave solution is obviously different from the plane wave solution of CGLE \eqref{GLplane_1}.
Similar to CGLE, the plane wave solutions may not be stable for all $\alpha$ and $\beta$, this can be studied by
Quasi-normal modes (QNMs) \cite{Horowitz2000,Kovtun2005}.
 \begin{figure}[t]
\centering
\includegraphics[width=1\linewidth]{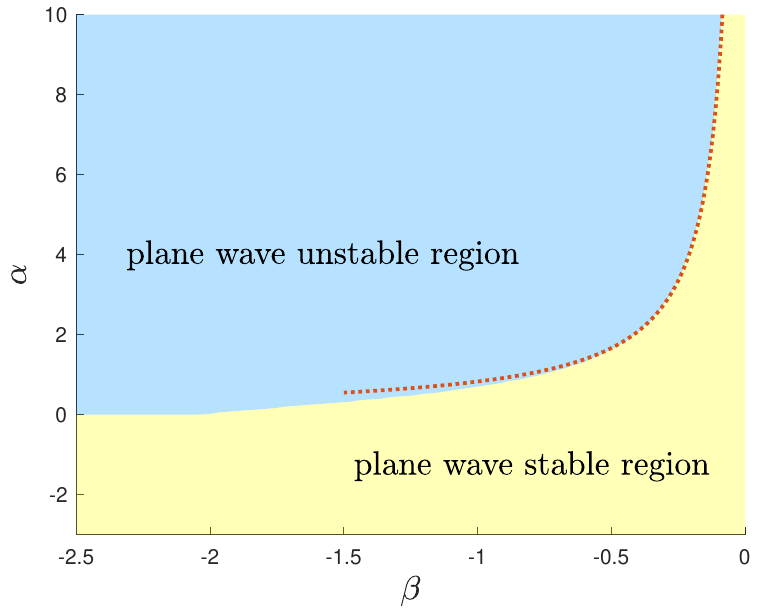}
\caption{Holographic BFN line and phase diagram of the holographic reaction-diffusion system. The red fitting line is $\alpha=(-0.83\pm 0.04)\beta^{-1.03+\pm0.02}$, which is different from the CGLE BFN line Eq.(\ref{glBF}).}\label{phasediagram}
\end{figure}

In the background of  plane wave solutions, the phase diagram of holographic reaction-diffusion systems can be obtained through the instability of the solutions. Like the usual linear instability analysis process, the plane wave solution with $q=0$, $\Psi=\Psi_{a}(z) e^{i \omega t}$ is
disturbed  by adding a small perturbation 
\begin{equation}
\Psi=(\Psi_a+\hat{\delta}_1(z) e^{\lambda t+ikx}+\hat{\delta}^*_2(z) e^{\lambda^* t-ikx}) e^{i \omega t}.
\label{addtur}
\end{equation}
Substitute Eq. \eqref{addtur} to the dynamic Eq. \eqref{EOM} we get the first order perturbation equations for $\delta_1$ and $\delta_2$
\begin{equation}
\left\{\begin{aligned}
& \hat{A} \hat{\delta}_1+\hat{C} \hat{\delta}_2=\lambda \hat{D}_t \hat{\delta}_1,\\
& \hat{A}^* \hat{\delta}_2+\hat{C}^* \hat{\delta}_1=\lambda \hat{D}_t^* \hat{\delta}_2.
\end{aligned}\right.
\label{QNMEOM}
\end{equation}
where $\hat{A}=z \left((1+i \alpha ) k^2 z-i \omega\right)+m^2+2 (1+i \beta ) \Psi_a \Psi_a^*+z \left(-z \partial_zf+f+2 i \omega z\right)\partial_z -z^2 f \partial_z^2$, $\hat{C}=(1+i \beta)\Psi_a^2$, $\hat{D}_t=z -2 z^2\partial_z$.
Solving the generalized eigenvalue Eq. \eqref{QNMEOM}, we can obtain a series of $\lambda$. Noting that only the eigenvalue $\lambda$ with the largest real part $Re(\lambda)$ is adopted as the growing rate for it determines the instability of holographic plane waves.  
The zero $q$ plane wave solution will be destroyed by the growing  perturbations if there are $Re(\lambda(k))>0$, corresponding to the linear unstable region in the phase diagram as shown in Fig.\ref{phasediagram}.
Four sample results of $Re(\lambda(k))$ for
different combinations of $\alpha$ and $\beta$ are given in Fig. \ref{QNM}.
If $Re(\lambda(k))\leq 0$, the plane wave solution is robust to the added perturbation, corresponding to the linear stable region in Fig.\ref{phasediagram}.
This is a holographic version of the BFN instability, where deviations from a periodic waveform solution are reinforced by nonlinearity, leading to the generation of spectral sidebands and the eventual breakup of the plane wave solution into a chaotic solution \cite{Benjamin1,Benjamin2}.
The holographic BFN line is temperature dependent, take $T=0.99T_c$ for example, it is exactly the BFN line \eqref{glBF} of CGLE (See appendix \ref{appB,appC} for a detailed review).

\textit{Spatiotemporal Chaos--}
Similar to CGLE \cite{Shraiman1992241}, when the holographic spatial extension system violates the BFN criterion, it exhibits irregular behavior in space and time: this phenomenon is commonly referred to as spatio-temporal chaos\cite{RevModPhys.65.851}. 
In particular beyond the BFN instability line but close to the critical line in Fig.\ref{phasediagram} exhibits so-called phase turbulence regime.
Phase turbulence is a state that $\mathcal{O}(x,t)=|\mathcal{O}|e^{i\theta}$ evolves irregularly, but with its modulus always fluctuates a bit near a constant value far from zero. For the phase $\theta$,  periodic boundary conditions force the winding number to be a constant of motion, fixed by the initial condition.
As can be
seen in Fig. \ref{chaos}, when $(\alpha,\beta)=(3,-0.65)$, this is a spatio-temporally chaotic state, the amplitude of order parameter never reaches zero
and remains saturated. Moreover, away from the BFN line, for example $(\alpha,\beta)=(3,-1.5)$ the system exhibits spatio-temporally disordered
regime called amplitude or defect turbulence. The behavior in this region is characterized by defects, where the order parameter vanishes (see Fig. \ref{chaos}) .  To obtain dynamics for the formation of chaos, we begin with a zero $\Psi$ plus
spatial noise of amplitude $h=10^{-3}$, admits the standard normal distribution. Sure,
we can also get the same results by beginning with the plane wave solution of the corresponding $\alpha$ and $\beta$ with spatial noise (not shown).  The linear instability analysis of the plane wave solution shown in Fig. \ref{QNM} confirmed that the plane solution will finally enter a chaotic state after a long-time
evolution, due to the exponential growth modes of finite $k$.

\begin{figure}
    \centering
    \includegraphics[width=1\linewidth]{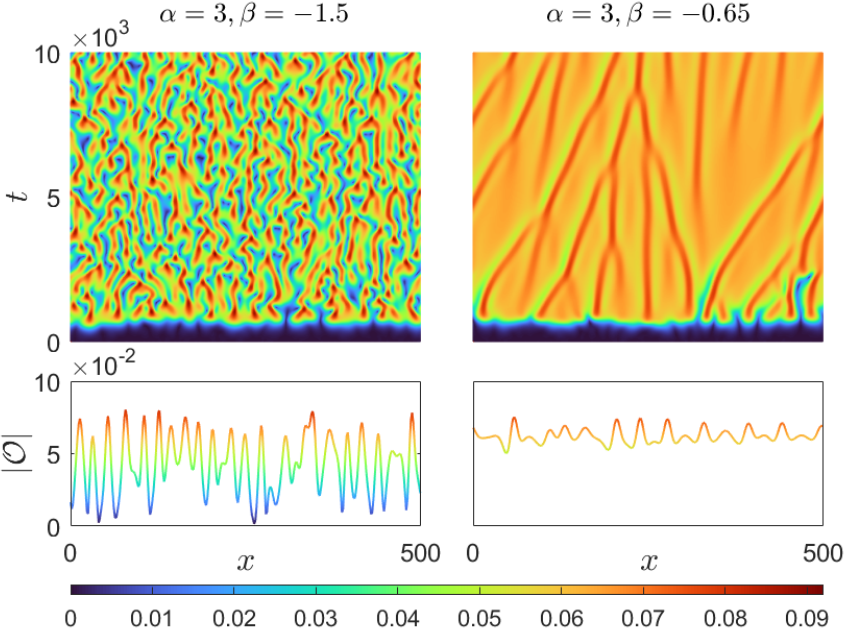}
    \caption{Configuration of $|\mathcal{O}(x,t)|$ for Defect turbulence (left) and phase turbulence (right) in the plane wave  unstable region. The column below shows $|\mathcal{O}(x)|$ at $t=10000$. }
    \label{chaos}
\end{figure}

\begin{figure}
    \centering
    \includegraphics[width=1\linewidth]{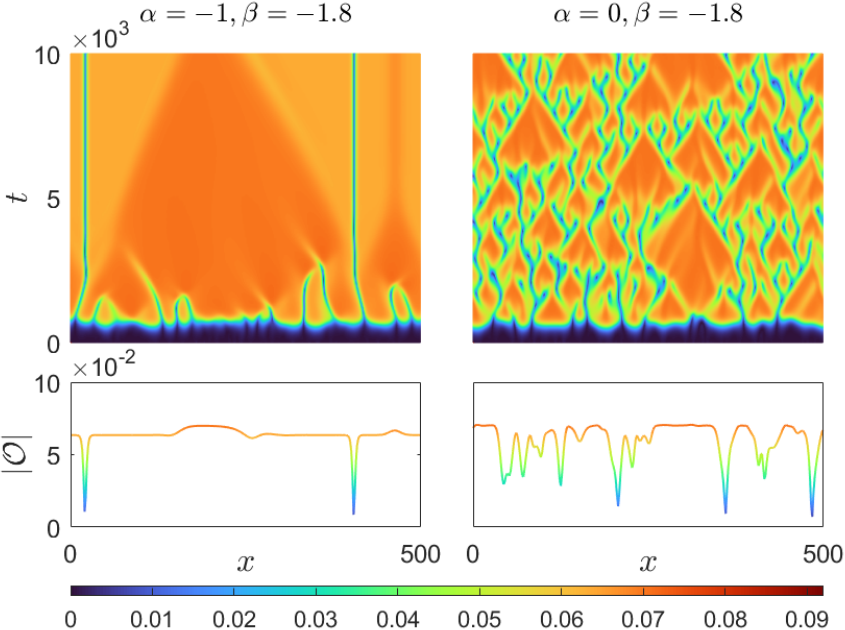}
    \caption{Configuration of $|\mathcal{O}(x,t)|$ for The moving hole-shock pair(left) and spatio-temporal intermittency  (right).The column below shows $|\mathcal{O}(x)|$ at $t=10000$.}
    \label{coherent}
\end{figure}
\textit{The spatio-tempora intermittency  and Coherent structures in the plane wave stable region--}
Even in the regime where plane waves are stable, where the perturbation with finite $k$ will exponential decay as shown in Fig. \ref{QNM},
the linear stability of the plane wave solution Eq. \eqref{planewave} can not exclude the existence
or coexistence of the other nontrivial solutions of Eq. \eqref{EOM}. Below the BFN line, plane waves attract most initial conditions. However, using a suitably large and localized initial condition, spatio-temporally intermittent states,  where defect chaos coexists with stable plane wave may appear as found in CGLE \cite{Chate_1994}.
After a rather short time evolution, a typical intermittency regime solution consists of localized structures, separating lager regions of almost constant amplitude which are patches of stable plane wave solution emerges, as shown in Fig. \ref{coherent} (right) when $(\alpha,\beta)=(0,-1.8)$,
with a configuration  very  similar to the observation in CGLE \cite{Chate_1994}.
 Fig. \ref{coherent} (left) shows another
 typical nontrivial solution called  sink solutions with Bekki-Nozaki holes \cite{19841581},
 observed for $(\alpha,\beta)=(-1,-1.8)$. The 
 strong experimental evidence for it  is given in the study of hydrothermal nonlinear waves\cite{Burguete1999}. 
 In this case, the spatial extension of the system is broken by irregular arrangements of stationary hole- and shock-like objects separated by turbulent dynamics. These structures asymptotically connect plane waves of different amplitude and wave number. Noting that the  solution illustrated in Fig. \ref{coherent} (left) are part of a family of solutions called coherent structures which are comprised of fixed spatial profiles that can vary through propagation and oscillation observed in CGLE \cite{VANSAARLOOS1992303}.

\textit{Discussion--}
We find that an extended holographic $U(1)$ symmetry broken 
theory shows pattern formation dynamics with 
many properties similar to CGLE, can relate to many true systems.
This is an enormous stretch on the application of 
Gauge/Gravity duality, shows that the holographic method has great potential to to study  numerous pattern formation dynamics besides the example we demonstrate. About the model we proposed, there are still many extensions needed to be studied: (i) Here we focus on the one spatial dimensional system closely related to the one dimensional CGLE,  in two or three dimensions there are many other nonequilibrium phenomena to be studied\cite{RevModPhys7499}; (ii) try to derive the effective boundary field theory of the holographic model for different temperatures; (iii) find the complete phase diagram to confirm if there exist other possible coherent structures had been observed in CGLE, including sinks, fronts, and shocks amongst others \cite{VANSAARLOOS1992303}.

\begin{figure}
    \centering
    \includegraphics[width=1\linewidth]{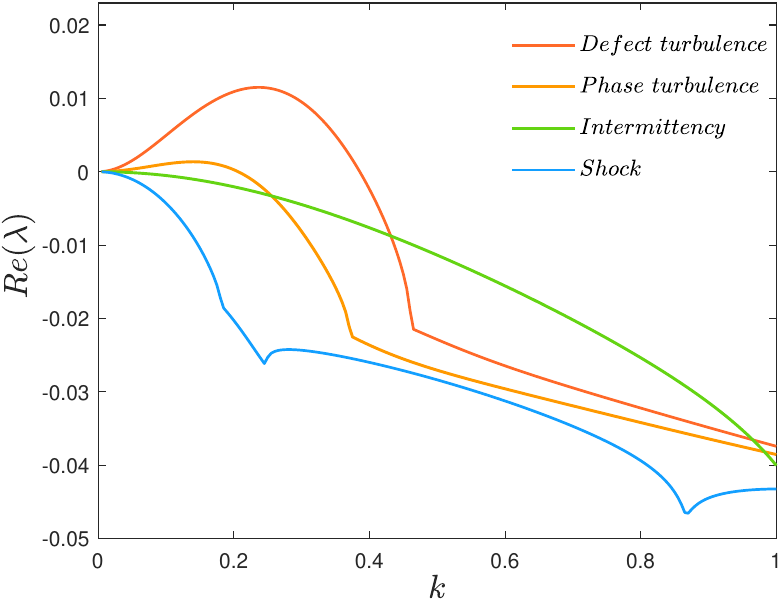}
    \caption{The real part of $\lambda$ (the growing rate of perturbation) versus $k$ in the background of plane wave solution, corresponding to the four situations shown in Fig. \ref{chaos} and Fig. \ref{coherent}, respectively are defect turbulence($(\alpha,\beta)=(3,-1.5)$), phase turbulence($(\alpha,\beta)=(3,-0.65)$), moving hole-shock($(\alpha,\beta)=(-1,-1.8)$), and Spatio-temporal intermittency($(\alpha,\beta)=(0,-1.8)$).}
    \label{QNM}
\end{figure}


{\it Acknowledgements.}
H.B. Z. acknowledges the support by the National Natural Science Foundation of China (under Grants No. 12275233)


\appendix
\maketitle

\section{Plan waves and Benjamin-Feir-Newell (BFN) instability of complex Ginzburg-Landau equation}\label{appA}
Substituting Eq. \eqref{GLperturbation} into Eq. \eqref{CGLE} one can obtain
\begin{equation}
\left\{\begin{aligned}
& A_c \delta_1+C_c \delta_2=\lambda \delta_1,\\
& A_c^* \delta_2+C_c^* \delta_1=\lambda  \delta_2.
\end{aligned}\right.
\end{equation}
This pair of equations can be organized into the eigenvalue equation of the matrix
\begin{equation}
\label{eigM}
\left(
\begin{array}{cc}
    A_c   & C_c \\
    C_c^* & A_c^* 
\end{array}
\right)
\left(
\begin{array}{c}
    \delta_1 \\ 
    \delta_2 
\end{array}
\right)
=\lambda
\left(
\begin{array}{c}
    \delta_1 \\ 
    \delta_2 
\end{array}
\right)
\end{equation}
where $A_c=1-k^2(1+i\alpha)-2(1+i\beta)-i\omega$ and $C_c=-(1+i\beta)$.
After solving the Eq. \eqref{eigM},  the expression of the eigenvalues $\lambda$ reads,
\begin{equation}
    \lambda_{\pm}=-1-k^2\pm\sqrt{1-k^4\alpha^2-2k^2\alpha\beta}
\end{equation}
The biggest eigenvalue  $\lambda_+$ which is called growth rate for it determines the instability of the plan waves.
When the real part of the $\lambda_+$ is less than zero, the plane wave is stable, otherwise it is unstable. 
By expanding $\lambda_+$ for small $k$ (the long-wavelength limit) one obtains
\begin{equation}
    \lambda_+=(-1-\alpha\beta)k^2+O(k^3).
\end{equation}
 $1+\alpha\beta>0$ is the Benjamin-Feir-Newell criterion, which only dependent on the parameters $\alpha$ and $\beta$. The outer boundary of the Benjamin-Feir-Newell instability is the so-called BFN line, 
\begin{equation}
    \label{glBF}
    \alpha=-\beta^{-1}
\end{equation}
above which all plane waves are unstable.
 \begin{figure}[t]
\centering
\includegraphics[width=1\linewidth]{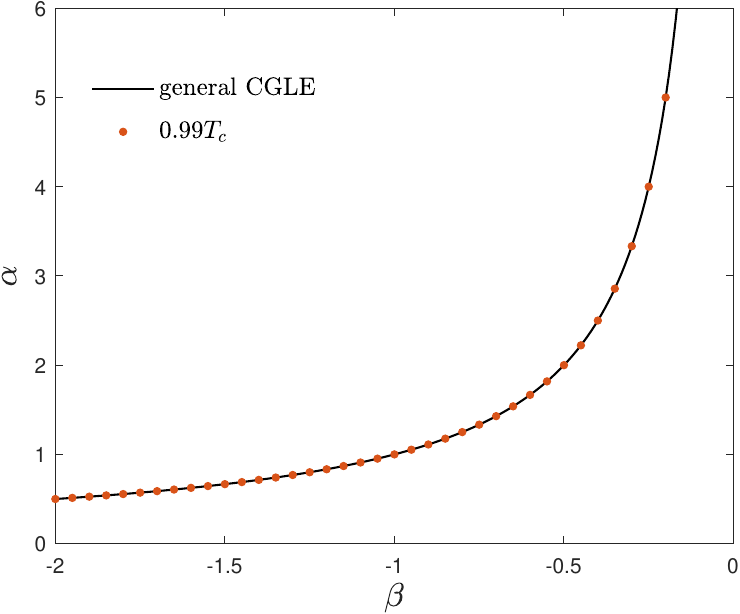}
\caption{Benjamin-Feir-Newell line of CGLE and holography. The result of the holographic BFN line at $T=0.99T_c$ is shown as red dots, which are fitted to $\alpha=-(0.993\pm0.003)\beta^{-1.004\pm0.004}$. The BFN line of the general CGLE \eqref{gCGLE} is shown in black line, which can be fitted to a perfect hyperbola $\alpha=-\beta^{-1}$. The perfect coincidence of these two curves indicates that the holographic model can be dual to the general CGLE \eqref{gCGLE}  near the critical temperature $T_c$.}\label{figS1:phasediagram}
\end{figure}

\begin{figure}
    \centering
    \includegraphics[width=1\linewidth]{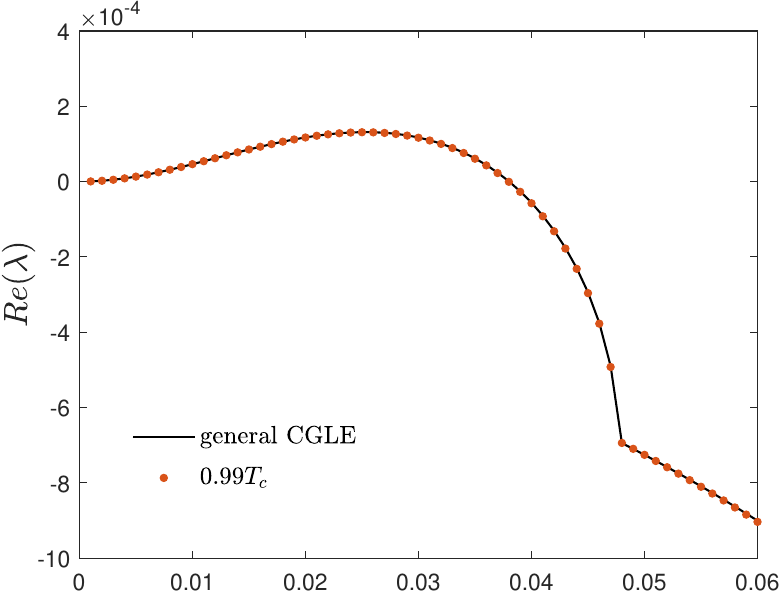}
    \caption{The real part of $\lambda$ (the growing rate of perturbation) versus $k$ in the background of plane wave solution, corresponding to the defect turbulence($(\alpha,\beta)=(3,-1.5)$). 
    The red dot line represents the results of holography at $T=0.99T_c$ while the black line represents the results of the general CGLE \eqref{gCGLE} with the same $\alpha$ and $\beta$. The perfect coincidence of these two curves supports that the holographic model can be dual to the general CGLE \eqref{gCGLE} near the critical temperature  $T_c$.
    }
    \label{figS2:QNM}
\end{figure}

\section{Plan waves and Benjamin-Feir-Newell (BFN) instability of Holographic pattern formation model near $T_c$}\label{appB}
In order to explore the properties of the holographic reaction-diffusion system near the critical temperature $T_c$, we fix the temperature to $T=0.99T_c$.
After solving the Eq. \eqref{EOMplane} by the Newton-Raphson iteration method at a range of parameters $(\alpha,\beta,q)$, the amplitude of the order parameter $\mathcal{O}_a$ and $\omega$ can be fitted, there are
\begin{equation}
\label{holoplane}
\mathcal{O}_a^2=\mathcal{O}_{0}^2-0.03143q^2, \quad
\omega=  -0.1586\alpha q^2-5.069\beta\mathcal{O}_a^2.
\end{equation}
Where $\mathcal{O}_0^2$ is defined at $q=\alpha=\beta=0$, specifically, $\mathcal{O}_{0}^2=6.442\times{10}^{-5}$.  
Obviously, the plane wave solution \eqref{holoplane} in holography near $T_c$ is very similar to that of CGLE Eq. ( \ref{GLplane_1}). 
In contrast,  the results of the holographic and CGLE  differ significantly at the temperature $T=0.29T_c$ that far from $T_c$ , indicating that the properties of plane waves are regulated by temperature.

Eq. \eqref{QNMEOM} can be organized into the eigenvalue equation of the matrix
\begin{equation}
\label{eigMholo}
\left(
\begin{array}{cc}
    \hat{A}   & \hat{C} \\
    \hat{C}^* & \hat{A}^* 
\end{array}
\right)
\left(
\begin{array}{c}
    \hat{\delta}_1 \\ 
    \hat{\delta}_2 
\end{array}
\right)
=\lambda
\left(
\begin{array}{cc}
    \hat{D}_t & 0 \\
    0   &   \hat{D}_t^*
\end{array}
\right)
\left(
\begin{array}{c}
    \hat{\delta}_1 \\ 
    \hat{\delta}_2 
\end{array}
\right)
\end{equation}
Solving this generalized eigenvalue Eq. \eqref{eigMholo}, we can obtain a series of $\lambda$. Noting that only the eigenvalue $\lambda$ with the largest real part $Re(\lambda)$ is adopted as the growing rate for it determines the instability of holographic plane waves.  If the growth rate $Re(\lambda(k))<0$, the zero $q$ plane wave solution is stable, otherwise it is unstable.  The  holographic version of Benjamin–Feir instability can be obtain when the growth rate $Re(\lambda(k))=0$.
The the numerical result of the holographic BFN line at $T=0.99T_c$ is shown in \ref{figS1:phasediagram}, which is fitted to 
\begin{equation}
\label{holoBF}
  \alpha=-0.993\beta^{-1.004}.  
\end{equation}
It is the same as the BFN line of the CGLE
Eq. (\ref{glBF}).

\section{Holographic pattern formation model near $T_c$ is  dual to complex Ginzburg Landau equation}\label{appC}
Except for the difference in coefficients, the numerical results Eq. \eqref{holoplane} and Eq.\eqref{holoBF} in holography at $T=0.99T_c$ are almost the same as \eqref{GLplane_1} and Eq.\eqref{glBF} of CGLE \eqref{CGLE}. In view of these evidences, the holographic pattern model near the critical temperature $T_c$ can qualitatively dual to the CGLE.
To be quantitative, we speculate that it is a  general CGLE but with several coefficients to be determined
\begin{equation}
\label{gCGLE}
\partial_t \mathcal{O}=a(1+i\alpha)\nabla^2\mathcal{O}-b(1+i\beta)|\mathcal{O}|^2\mathcal{O} +c \mathcal{O},
\end{equation}
where  $(a,b,c)$ are the three real undetermined coefficients. 
The plane wave solutions of Eq. \eqref{gCGLE} are  $\mathcal{O}= \mathcal{O}_a e^{i \omega t+ i q x}$,
where
\begin{equation}
\label{gGLplane}
\mathcal{O}_a^2 =\frac{c}{b} -\frac{a}{b}q^2, \quad \omega = -a\alpha q^2 - b\beta \mathcal{O}_a^2.
\end{equation}
By comparing Eq. \eqref{GLplane_1}  and Eq. \eqref{gGLplane}, we can obtain four equations about the three coefficients.
By using any three of the four equation, 
can $(a,b,c)$ be accurately calculated, 
\begin{equation}
\label{abc}
(a,b,c)=(0.1586,5.069,3.255\times{10}^{-4}).
\end{equation}
Then we found results \eqref{abc} can be accurately verified by the remaining equation.
This is also an evidence that this holographic model can be well dual to the general CGLE \eqref{gCGLE}.

To obtain stronger evidence, let us discuss the instability of plane wave Eq. \eqref{gGLplane} of the general CGLE \eqref{gCGLE}.
The perturbation is introduced as 
\begin{equation}
\label{gGLperturbation}
\mathcal{O}=[\mathcal{O}_a+\delta_{1} e^{\lambda t+ikx}+\delta^*_{2} e^{\lambda^* t-ikx}] e^{i \omega t},
\end{equation} 
where the scalars $\delta_{1}$ and $\delta^*_{2}$
 denote the amplitudes of the small perturbations.
Then substituting Eq. \eqref{gGLperturbation} into Eq. \eqref{gCGLE} one can obtain
the eigenvalue equation 
\begin{equation}
\label{geigM}
\left(
\begin{array}{cc}
    A   & C \\
    C^* & A^* 
\end{array}
\right)
\left(
\begin{array}{c}
    \delta_1 \\ 
    \delta_2 
\end{array}
\right)
=\lambda
\left(
\begin{array}{c}
    \delta_1 \\ 
    \delta_2 
\end{array}
\right)
\end{equation}
where $A=c-ak^2(1+i\alpha)-2(1+i\beta)c-ic\omega$ and $C=-c(1+i\beta)$.
After solving the Eq. \eqref{geigM},  the expression of growth rate $\lambda$ reads,
\begin{equation}
\label{g_eig}
    \lambda_{}=-c-ak^2+\sqrt{c^2-a^2k^4\alpha^2-2ack^2\alpha\beta}
\end{equation}
By expanding it for small $k$ (the long-wavelength limit) one obtains
\begin{equation}
\label{gBF}
    \lambda_{}=(-c-c\alpha\beta)k^2+O(k^3).
\end{equation}
From Eq. \eqref{gBF}, it is easy to find  that the BFN line of the general CGLE is also $\alpha=-\beta^{-1}$ which has nothing to do with parameters $(a,b,c)$. The numerical result of the holographic BFN line (Eq. \eqref{holoBF}) in \ref{figS1:phasediagram} perfectly verifies this.
What is even more interesting is that, as shown in the Fig. \ref{figS2:QNM}, the direct numerical calculation of $Re(\lambda)$ using \eqref{eigMholo} is in perfect agreement with the result of Eq. \eqref{g_eig} under the the fitted parameters \eqref{abc}.

To sum up, all the results  support that the holographic reaction-diffusion system near $T_c$ is dual to the general CGLE \eqref{gCGLE} very well.

\bibliography{holoPF_v2}

\end{document}